# The role of vibrational temperature variations in a pulsed dielectric barrier discharge plasma device

Fellype do Nascimento[1*], Stanislav Moshkalev[1], and Munemasa Machida[2],

[1]*Center for Semiconductor Components – State University of Campinas, C. P. 6101, Campinas, CEP 13083-870, Brazil*

[2] *Instituto de Física "Gleb Wataghin" – State University of Campinas, Campinas, CEP 13083-859, Brazil*

**Abstract**

The use of dielectric barrier discharge (DBD) plasmas has become a practical way to carry out surface treatment and one seeks to do it in a more efficient way, which requires to have control of the plasma parameters like rotational and vibrational temperatures ($T_{rot}$ and $T_{vib}$). Since the $T_{vib}$ of an atmospheric pressure plasma jet is an important parameter related to improvement of surface treatments, in this work it was analyzed two methods to increase the values of $T_{vib}$ in the DBD plasma. One of the methods is to reduce the exit size of the DBD reactor, and the other is to change the gas flow rate. Explanations for the reasons that lead to the increment of the vibrational temperature are proposed in both cases.

**Introduction**

Cold plasmas produced at atmospheric pressure and in open environments have received considerable attention in recent years due to their versatility, easiness of operation and low cost compared to plasmas produced in vacuum environment [Moon2004,Lu2014]. These plasmas are characterized by low temperature that is especially important for the modification and/or activation of surfaces of soft materials like polymers or biological tissues, without damaging them [Masoud2005,Rajasekaran2012,Bashir2014,Machida2015]. A plasma generated using dielectric barrier discharge (DBD) is one kind of atmospheric pressure plasma in which the discharge is produced between two electrodes with at least one of them covered with a dielectric material (glass or ceramic in most cases) [Kogelschatz1997,Corke2010].

Most DBD devices that operate at atmospheric pressure and in open environments can use many different gases as the working gas [Nascimento2016, Masoud2005, Yuji2007,Keller2012, Muller2013]. Depending on the working gas, the resulting plasma presents different values for parameters like vibrational and rotational temperatures, electron density, delivered power, and so on [Bibinov2001, Masoud2005, Faure1998, Moon2004, Yuji2007, Walsh2010], and the variation of these plasma parameters can affect surface treatment processes. We can also change the plasma parameters for a given working gas if the operation conditions are changed, like varying the gas flow rate, the gas pressure, and/or the voltage applied on the electrodes.

---

*E-mail: fellype@gmail.com

The variation of plasma rotational and vibrational temperatures with operation parameters like gas flow rate and gas pressure were subject of study by some authors [Masoud2005, Yuji2007, Moon2002, Wang2007], but in none of these studies a theoretical model was proposed to explain the results. Yuji *et al* [Yuji2007] showed that the values of WCA in poly(ethylene terephthalate) (PET) decrease when surface treatments using plasmas with higher rotational temperature are performed, but that study did not take in to account the effects of $T_{vib}$ values, which has been shown to be an important plasma parameter on the improvement of surface treatment processes [Nascimento2017].

In this work we analyze two ways to increase or reduce the vibrational temperature of the DBD plasma. One is by reducing the exit size of the DBD reactor, or, in other words, putting a smaller amount of the plasma inside the reactor in contact with the surrounding air. And the other is changing the gas flow rate used to produce the plasma. In the first case we verified that the larger the exit size, the lower the vibrational temperature. But in the last case the behavior of vibrational temperature as a function of gas flow rate depends on the working gas used to produce the plasma. We also propose explanations for the reasons that lead to the increment or reduction of the vibrational temperature in both cases.

**Theoretical background**

*- Vibrational excitation*

In DBD plasmas there are some processes that can change the number density of $N_2$ molecules in vibrationally excited states, increasing the vibrational energy of the gas: collisions with electrons, collisions with metastable atoms and ions, energy transfer by collisions between vibrationally excited molecules, and conversion of vibrational energy into translational and/or rotational motion of the molecules [Märk1971, Lindinger1981, Plain1984, Piper1987, Brandenburg2005, Peñano2012].

In Fig. 1 we can see that metastable helium He(2 $^3$S) and argon ion $Ar^+(^3P_{3/2})$ have enough energy to ionize $N_2$ molecules to states $N_2^+$(A $^2\Pi_u$ and B $^2\Sigma_u^+$) for He, and $N_2^+$(X $^2\Sigma_g^+$) for $Ar^+$. The $N_2^+$ ions and argon metastable states Ar($^3P_2$ and $^3P_0$) can collide with $N_2$ and produce excited states $N_2$(C $^3\Pi_u$ and B $^3\Pi_g$) and metastable states $N_2$(A $^3\Sigma_u^+$ and $a'$ $^1\Sigma_u^-$). Collisions between two metastable $N_2$ and between a $N_2^+$ with a ground state $N_2$(X $^1\Sigma_g^+$) can also produce excited $N_2$ ($N_2^*$) molecules. Table 1 summarizes the most important of the possible reactions that can produce excited $N_2$ molecules and their rate coefficients at room temperature. In Table 1, penning ionization of $N_2$ by argon metastable ($Ar^M$) and by $N_2$(A $^3\Sigma_u^+$) are neglected because the energies of $Ar^M$ and $N_2$(A $^3\Sigma_u^+$) are considerably lower than the ionization energy of $N_2$, as can be seem in Fig.1. Penning ionization due to collisions between metastables $N_2(A\ ^3\Sigma_u^+)$ and $N_2(a'\ ^1\Sigma_u^-)$ are neglected because its contribution is very small when compared with other processes that can occur with these metastable [Guerra2003]. Penning excitation of $N_2$ by helium metastable ($He^M$) can also be neglected because the energy of $He^M$ is considerably higher than the energy of the $N_2(C\ ^3\Pi_u)$ state. The reaction $e + N_2 \rightarrow 2e + N_2^+$ is neglected because its cross section is very small for the electron energies that can be found in a DBD plasma [Märk1975].

Neutral $N_2$ molecules are vibrationally excited by electron impact at a rate of $4\times10^{-9}$ cm$^3$ s$^{-1}$, for an electron temperature of 1 eV [Fridman2008]. Additionally, an electronically excited $N_2$ molecule ($N_2^*$), produced by any of the reactions shown in Table 1, lies on a vibrationally excited state v', since metastables, ions, and excited molecules are able to perform vibrational excitation on the $N_2$ molecules [Tozeau1978,Belikov1997].

Observing the data in Table 1, we notice that reaction number 10 is one of the most important source of $N_2^*$ because it has the highest reaction coefficient related to $N_2^*$ production in plasmas at room temperature, and, since it depends on reactions 6 or 9, we see that there are two ways to produce a plasma with a high vibrational energy in an open environment. One is using a working gas that is able to produce a large amount of $N_2^+$ ions. Comparing the energy levels of $N_2^+$ ions with those of $He(2\ ^3S)$ and $Ar^+(^3P_{3/2})$ (see Fig. 1), we can infer that the helium gas is the best choice in order to produce a plasma with a high degree of vibrational excitation. The other way is to use a working gas that is able to produce a large amount of $N_2^M$ metastables, which can be done using pure nitrogen or argon as working gases, as suggested by the energy level diagram shown in Fig. 1.

**Table 1:** Most important of the possible reactions that can produce excited $N_2$ molecules and their rate coefficients at room temperature.

| Number | Reaction [a] | Rate coefficient | Reference |
|---|---|---|---|
| 1 | $Ar^M + N_2 \rightarrow Ar + N_2^*$ | $2.9\times10^{-11}$ (cm$^3$ s$^{-1}$) | Dyatko2010 |
| 2 | $Ar^+ + N_2 \rightarrow Ar + N_2^+$ | $0.1\text{-}6.6\times10^{-11}$ (cm$^3$ s$^{-1}$) | Lindinger1981 |
| 3 | $He^M + N_2 \rightarrow He + (N_2^{+*}, N_2^+)$ | $0.63\text{-}1\times10^{-10}$ (cm$^3$ s$^{-1}$) | Märk1971 |
| 4 | $N_2^* + N_2 \rightarrow N_2 + N_2^*$ | $1.9\times10^{-13}$ (cm$^3$ s$^{-1}$) | Piper1987 |
| 5 | $N_2^M + N_2^M \rightarrow N_2^* + N_2$ | $1.5\text{-}2.6\times10^{-10}$ (cm$^3$ s$^{-1}$) | Bibinov2001b,Hays1973 |
| 6 | $N_2^M + N_2^M \rightarrow N_4^+ + e$ | $1\text{-}5\times10^{-11}$ (cm$^3$ s$^{-1}$) | Guerra1997 |
| 7 | $N_2^{+*} + N_2 \rightarrow$ products | $3.7\text{-}8.2\times10^{-10}$ (cm$^3$ s$^{-1}$) | Plain1984,Bibinov2001,Anicich1993 |
| 8 | $e + N_2 \rightarrow e + N_2^*$ | $3.4\times10^{-10}$ (cm$^3$ s$^{-1}$) | Brandenburg2005 |
| 9 | $N_2^+ + N_2 + N_2 \rightarrow N_4^+ + N_2$ | $6.8\times10^{-29}$ (cm$^6$ s$^{-1}$) | Peñano2012 |
| 10 | $N_4^+ + e \rightarrow N_2^* + N_2$ | $3.2\times10^{-7}$ (cm$^3$ s$^{-1}$) [b] | Peñano2012 |

[a] M = metastable state; * = electronically excited; + = ionized.
[b] for an electron temperature of 1 eV.

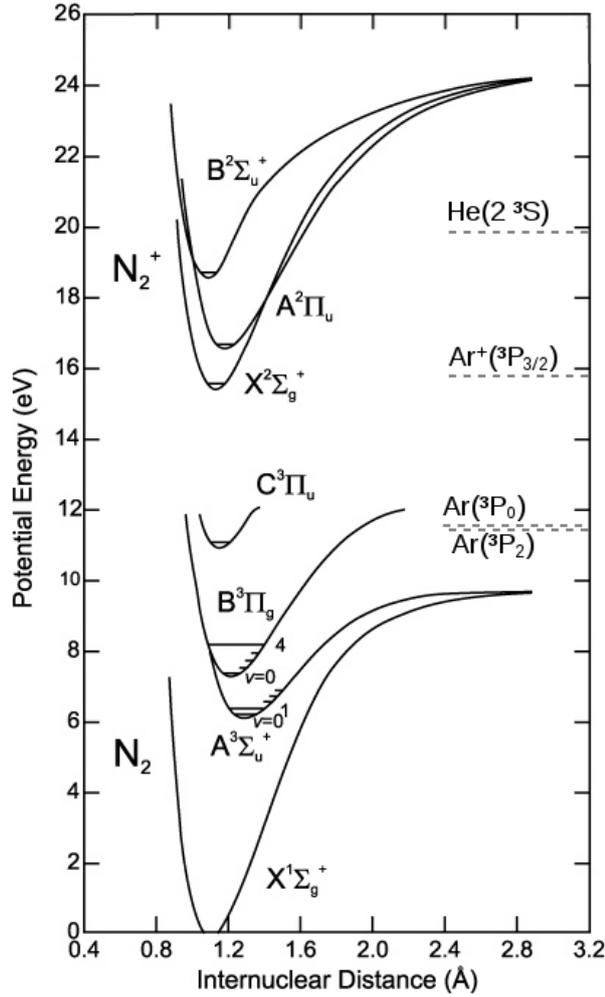

**Figure 1:** Energy level diagram for molecular nitrogen, with references to the energies of He(2 $^3$S), Ar($^3$P$_2$,$^3$P$_0$), and Ar$^+$($^3$P$_{3/2}$) states (figure adapted from [Ono2009])

 *- Vibrational temperature*

The total energy E of an excited N$_2$ molecule is

$$E = E_e + E_{trans} + E_{vib} + E_{rot} \tag{1}$$

where $E_e$ is the electron energy, $E_{trans}$ is the translational energy, $E_{vib}$ is the vibrational energy, and $E_{rot}$ is the rotational energy.

Assuming a Boltzmann distribution for the number density of molecules in a rotational-vibrational excited state $N_{v'J'}$, we can write [Zheng-De2002,Wang2009]:

$$N_{v'J'} = N_0 \alpha(v', J') \exp(-E_{v'}/k_B T_{vib}) \exp(-E_{J'}/k_B T_{rot}) \tag{2}$$

where $N_0$ is the number density of N$_2$ molecules that are not excited (neutral), α(v',J') is a constant of the state v'-J', $E_{v'}$ is the energy of the vibrational state v', $E_{J'}$ is the energy of the rotational state J', $k_B$ is the Boltzmann constant, and $T_{vib}$ and $T_{rot}$ are the vibrational and rotational temperatures, respectively. For non-equilibrium plasmas the rotational temperature is known to be approximately equal to the gas temperature ($T_{rot} \approx T_{gas}$) [Motret2000,Moon2003,Bruggeman2014].

The N$_2$ molecules that enters the plasma have a partial pressure $P_0$ and are subjected to the gas temperature in the plasma. When working on atmospheric pressure and in an open environment we can assume that the $N_2$ gas is an ideal one, which allows us to use the ideal gas equation, $P_0 = N_0 k_B T_{gas}$, together with the approximation $T_{gas} \approx T_{rot}$ in order to write the vibrational temperature in terms of $P_0$, $N_{v'J'}$, and $T_{rot}$ after substitute all this in Eq. (2) resulting in:

$$T_{vib} = \frac{E_{v'}}{k_B} \left[ \ln\left(\frac{P_0 \alpha(v', J')}{N_{v'J'} k_B T_{rot}}\right) - \frac{E_{J'}}{k_B T_{rot}} \right]^{-1} \quad (3)$$

From Eq. (3) we can see that $T_{vib}$ decreases if $P_0$ is increased, and increases if $N_{v'J'}$ and/or $T_{rot}$ are increased. If $P_0$ is increased, the mean free path for collisions between $N_2$ molecules in the plasma becomes smaller, which reduces the vibrational energy transferred in each collision, and then $T_{vib}$ is reduced. When neutral $N_2$ molecules enter the plasma they start to collide with energetic electrons and excited $N_2$ molecules, resulting in a larger number of vibrationally excited molecules and, consequently, in a higher $T_{vib}$. The gas temperature acts giving more kinetic energy to the molecules in the plasma, then the higher $T_{gas}$, the higher $T_{vib}$.

**Experimental procedure**

The experimental setup used to measure both $T_{vib}$ and $T_{rot}$ of DBD plasmas are shown in Fig. 2. The basic functioning of the device is as follows: a continuous gas flow is injected inside the poly(vinyl chloride) (PVC) tube and high-voltage pulses are applied to the electrode inside the glass tube. A primary plasma discharge is formed in the region between the glass tube and the PVC tube producing a plasma jet leaving the tube exit.

The pulsed voltage applied to the electrode had peak values of 15 kV when Ar and He gases were used and 30 kV when the $N_2$ gas was used. The frequency of plasma pulses was 60 Hz in all cases. The distance **d** between the end of the PVC tube and the surface of the dielectric plate was kept constant and equals to 5 mm in all experiments performed in this work.

In order to obtain the temperatures as a function of the diameter ø of the PVC tube exit, the changeable part of the device (dashed part in Fig. 2) is replaced by another one with a different exit diameter.

Both $T_{vib}$ and $T_{rot}$ were determined using the SpecAir software [refSpecAir] and measurements of spectral emissions were performed using an Andor 303i spectrometer equipped with an iStar DH720 iCCD detector and a 1200 lines/mm grating. The light emitted by the plasmas was collected with a lens and transported to the spectrometer through an optical fiber. The vibrational temperature of $N_2$ molecules was determined comparing a simulated spectrum that best fits the experimental one. In order to obtain rotational and vibrational temperatures of $N_2$ molecules, we use emission lines from the second positive system, C $^3\Pi_u$, v' → B $^3\Pi_g$, v'', where C $^3\Pi_u$ and B

$^3\Pi_g$ denote electron configuration, and v' and v" denote the vibrational state. Each vibrational band comes from a transition of different vibrational state (v' → v").

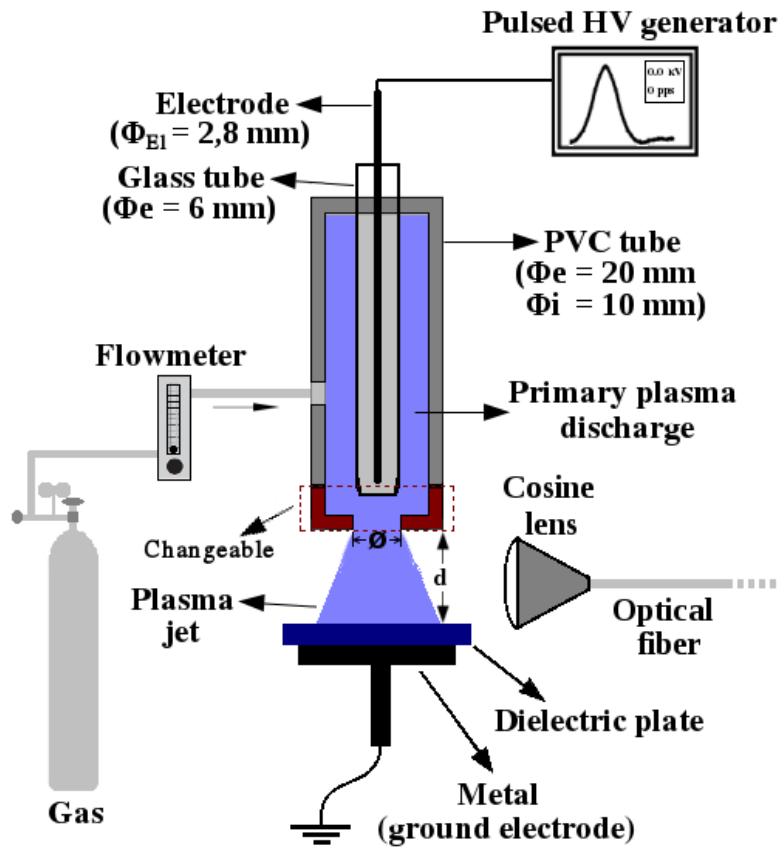

**Figure 2:** DBD scheme used for temperature measurements and for plasma treatment of PDMS samples. Φe and Φi refer to external and internal diameters, respectively. The elements are out of scale.

**Results and discussion**

In Figure 3(a) are shown curves of the variation of $T_{vib}$ as a function of the diameter of the PVC tube exit obtained for plasmas produced using Ar, He and $N_2$ as the working gases and Fig. 3(b) shows the respective curves obtained for $T_{rot}$.

As we can see in Fig. 3(a), the $T_{vib}$ values for Ar and He plasmas decrease if the diameter of the PVC tube exit (ø) is increased. The larger variation occurs for the He plasma. The $T_{vib}$ values for $N_2$ plasma does not change significantly when ø is increased. From Fig. 3(b) we can see that the $T_{rot}$ values does not change significantly with the size of the PVC opening when He and $N_2$ gases are used, but tends to decrease when Ar is used.

When ø is increased the gas pressure inside the PVC tube is reduced and the surrounding air tends to have more contact with the primary plasma inside the PVC tube. Then we have an increase in the amount of $N_2$ molecules, that come from the air, in the plasma, which increases the number density of molecules in all vibrationally excited states and results in an increment in the vibrational temperature when Ar and He are used to generate plasmas, which is in agreement with theory. The reduction of the pressure of Ar and He gases inside the PVC tube allows an increase in the partial pressure of $N_2$ molecules coming from air, and it also contributes with the reduction of the $T_{vib}$ values.

When the $N_2$ gas is used, the reduction in the gas pressure inside the PVC tube due to the increase on the size of the exit opening is accompanied by a reduction of the number of $N_2$ molecules that are compressed in the PVC tube, which results in a lower number of molecules in higher vibrational states, and then the $T_{vib}$ values do not change significantly.

In another experiment with an atmospheric pressure plasma jet, Yuji *et al* [Yuji2007] increased the amount of nitrogen gas in the plasma increasing the $N_2$ gas flow rate in an Ar/$N_2$ gas mixture, keeping the Ar flow rate constant. As a result, they obtained a curve which showed that the $T_{vib}$ values increases as the $N_2$ gas flow rate increases. Once the neutral molecules enter the plasma they become excited and $T_{vib}$ is increased, and this confirms the dependence of $T_{vib}$ with $N_{v'J'}$ predicted by Eq. (3).

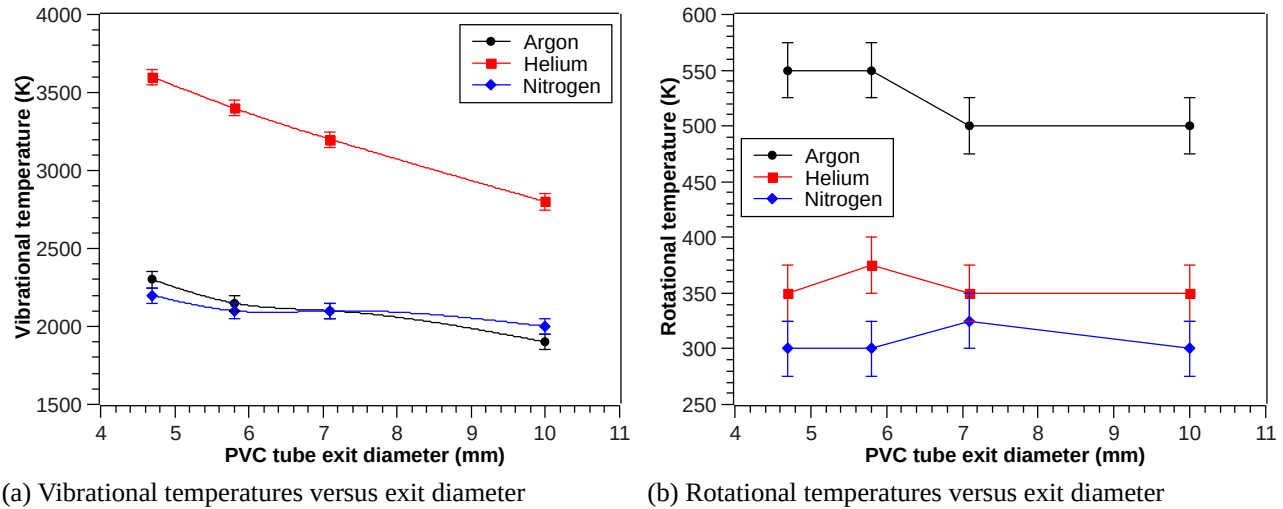

(a) Vibrational temperatures versus exit diameter   (b) Rotational temperatures versus exit diameter

**Figure 3:** Vibrational (a) and rotational (b) temperatures in function of the diameter of the PVC tube exit using a gas flow rate of 4 L/min.

Figure 4(a) shows curves of the variation of $T_{vib}$ as a function of the gas flow rate obtained for plasmas generated using Ar, He and $N_2$ as working gases and Fig. 4(b) shows the respective curves obtained for $T_{rot}$.

From Fig. 4(a) we can see that the $T_{vib}$ values for He and $N_2$ plasmas decrease if the gas flow rate is increased but those for Ar plasma do the opposite. In Fig. 4(b) we can see that the increment of the gas flow rate does not affect significantly the $T_{rot}$ values for He and $N_2$ plasmas but results in increment of this parameter in the case when Ar is used.

There is a different explanation for the behavior of $T_{vib}$ in relation to the gas flow rate for each gas used. For the cases of operation with Ar and He gases, if the gas flow rate is increased, the total gas pressure inside the PVC tube increases, but the partial pressure of $N_2$ molecules decreases, and the number of $N_2$ molecules is reduced, reducing also the number of $N_2$ molecules vibrationally excited. For the case of operation with $N_2$ gas, both the gas pressure and the number of $N_2$ molecules vibrationally excited increases with the gas flow rate. The reduction of the $T_{vib}$ values with the gas flow rate for He plasmas should be related with the reduction of the number of $N_2$ molecules vibrationally excited. The reduction of $T_{vib}$ values for $N_2$ plasmas may be related with the increment in gas pressure. Both cases are in agreement with theory.

For the Ar plasma, the increment in $T_{vib}$ values as a function of the gas flow rate is related with the increment in the respective $T_{rot}$ values, and this is in agreement with predictions given by Eq. (3). There is also a variation of the gas pressure inside the PVC tube with the gas flow rate, but it is lower than the variation of $T_{rot}$ values.

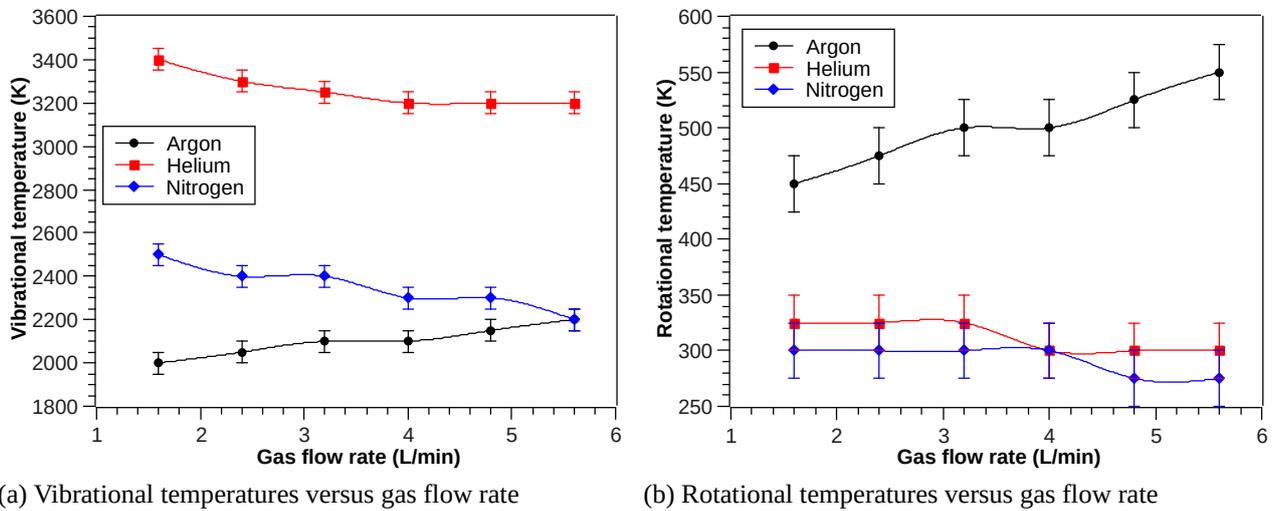

(a) Vibrational temperatures versus gas flow rate  (b) Rotational temperatures versus gas flow rate

**Figure 4:** Vibrational (a) and rotational (b) temperatures in function of the gas flow rate, which is proportional to the gas pressure, using a PVC tube exit of 7.1 mm diameter.

The behavior of the $T_{vib}$ as a function of the gas pressure for He and $N_2$ plasmas are in agreement with that reported by Masoud *et al* [Masoud2005], in which the $T_{vib}$ values decrease when the gas pressure increases.

The increment in $T_{rot}$ (or $T_{gas}$) values when the gas flow rate is increased in the case where Ar is used as the working gas was also observed by Moon *et al* [Moon2002] using an atmospheric microwave-induced plasma. This behavior is in contrast to the expected one because, if other parameters like the applied voltage and pulse frequency are kept constant, an increase in the gas flow rate usually tends to cool down a discharge due to the increase in the number of particles in the plasma. Some authors reported an increment in the electron density by increasing the gas flow rate when working with Ar gas [Moon2002,Humud2015]. This fact combined with the higher electrical conductivity of Ar gas may be the reason for the increment in $T_{rot}$, since a larger number of collisions between electrons and nitrogen molecules can increase the $T_{rot}$ and warm up the plasma.

As we can see in Figs. 3(a) and 4(a), the higher values of $T_{vib}$ were obtained using helium as the working gas, and these values are considerably higher than in the cases in which argon or nitrogen is used. In order to investigate the reasons for such differences, we obtained the emission spectrum of the plasmas formed using the three gases. The results are shown in Fig. 5.

From Fig. 5, we notice that only the plasma made using the He gas presents detectable molecular line emissions from $N_2^+$ ion ($N_2$ II) at 391.4 and 427.8 nm. If we compare the intensities of the molecular line emissions from $N_2$ molecule ($N_2$ I), we observe that they are higher in the He plasma. Then, considering the higher $T_{vib}$ values obtained with He gas, we observe that the penning ionization process by metastable helium plays an important role in the increment of vibrational excitation of $N_2$ particles in the plasma. From Fig. 5 we can also deduce that the penning ionization processes associated with $N_2$ and Ar metastables are negligible in DBD plasmas at atmospheric pressure, since there are no detectable line emissions from the $N_2^+$ ion in the emission spectra of the plasmas made with these gases. The lower $T_{vib}$ values obtained for operation with Ar and $N_2$ gases indicate that reaction 6 (see Table 1) does not contribute significantly to the generation of $N_4^+$ species, which are used in the reaction 10 to generate excited $N_2$ molecules, in DBD plasmas.

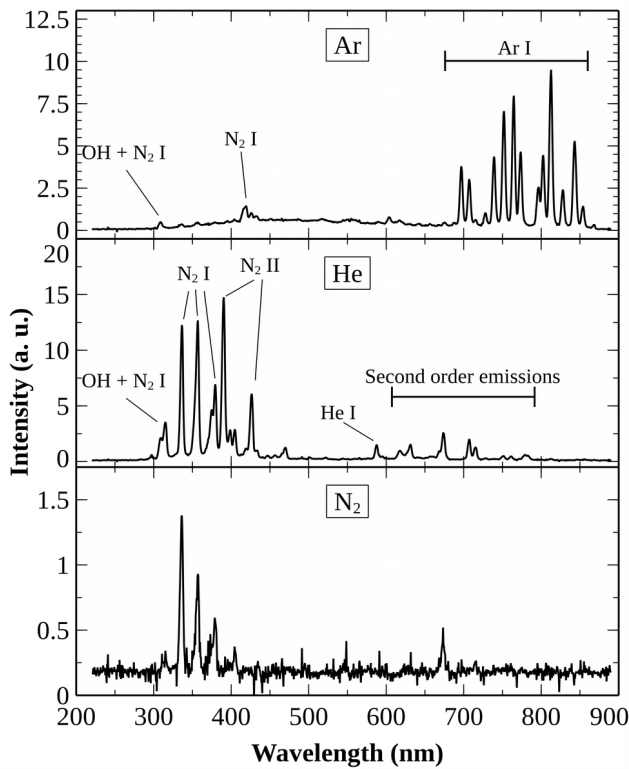

**Figure 5:** Emission spectra for plasmas using argon, helium and nitrogen as working gases..

**Conclusions**

According to the curves shown in Fig. 3(a) where the $T_{vib}$ values decreases with increasing diameter of the exit of the device, if the plasma inside the reactor has less contact with the environment due to the smaller size of device exit, it results in higher $T_{vib}$ values. This happens regardless of which working gas is used to produce the plasma and, according to Eq. (3), that behavior is associated to the reduction of the gas pressure inside the reactor due do the increment of the exit diameter, which allows more $N_2$ molecules from the environment to have contact with the primary plasma discharge.

As the curves show in Fig. 4-a, the behavior of the $T_{vib}$ values as a function of the gas flow rate are similar – higher the flow rate, lower the $T_{vib}$ values – when He or $N_2$ are used to produce the plasmas, but it increases that parameter when Ar is used. All the $T_{vib}$ results obtained by variation of the gas flow are in agreement with predictions given by Eq. (3), which confirms the validity of the model.

The use of helium is essential to achieve the highest $T_{vib}$ values in DBD plasmas, because only it has metastable states with energy high enough to ionize a large amount of $N_2$ molecules, and the $N_2^+$ ions looks to be the primarily responsible for the reactions that lead to increased $T_{vib}$ in the plasma. It is not possible to reach higher values of $T_{vib}$ using argon and nitrogen gases because the production of $N_2^+$ ions is negligible in these cases.

**References**

[Moon2004] SY Moon, W Choe, BK Kang, "A uniform glow discharge plasma source at atmospheric pressure", Appl. Phys. Lett. **84** (2004) 188

[Lu2014] X Lu, G V Naidis, M Laroussi, K Ostrikov, "Guided ionization waves: Theory and experiments", Phys. Rep. **540** (2014) 123

[Masoud2005] N Masoud, K Martus, M Figus, K Becker, "Rotational and Vibrational Temperature Measurements in a High-Pressure Cylindrical Dielectric Barrier Discharge (C-DBD)", Contrib. Plasma Phys. **45** (2005) 30

[Rajasekaran2012] P Rajasekaran, N Bibinov, P Awakowicz, "Quantitative characterization of a dielectric barrier discharge in air applying non-calibrated spectrometer, current measurement and numerical simulation", Meas. Sci. Technol. **23** (2012) 085605 (8pp)

[Bashir2014] M Bashir, J M Rees, S Bashir, W B Zimmerman, "Characterization of atmospheric pressure microplasma produced from argon and a mixture of argon–ethylenediamine", Phys. Lett. A **378** (2014) 2395

[Machida2015] M Machida, "Ferrite Loaded DBD Plasma Device", Braz. J. Phys. **45** (2015) 132

[Kogelschatz1997] U Kogelschatz, B Eliasson, W Egli, "Dielectric-Barrier Discharges. Principle and Applications", J. Phys. IV **7 (C4)** (1997) C4-47

[Corke2010] T C Corke, C L Enloe, S P Wilkinson, "Dielectric Barrier Discharge Plasma Actuators for Flow Control", Annu. Rev. Fluid Mech. **42** (2010) 505

[Bodas2006] D Bodas, C Khan-Malek, "Formation of more stable hydrophilic surfaces of PDMS by plasma and chemical treatments", Microelectron. Eng. **83** (2006) 1277

[Nascimento2016] F Nascimento, S Parada, S Moshkalev, M Machida "Plasma treatment of poly(dimethylsiloxane) surfaces using a compact atmospheric pressure dielectric barrier discharge device for adhesion improvement", Jpn. J. Appl. Phys. **55** (2016) 021602

[Yuji2007] T Yuji, Y Suzaki, T Yamawaki, H S Akaue, H A Katsuka, "Experimental Study of Temperatures of Atmospheric-Pressure Nonequilibrium Ar/$N_2$ Plasma Jets and Poly(ethylene terephtalate)-Surface Processing", Jpn. J. Appl. Phys. **46** (2007) 795

[Nascimento2017] F Nascimento, M Machida, M A Canesqui and S A Moshkalev, "Comparison Between Conventional and Transferred DBD Plasma Jets for Processing of PDMS Surfaces", IEEE Trans. Plasma Sci. **45** (2017) 346

[Keller2012] S Keller, P Rajasekaran, N Bibinov, and P Awakowicz, "Characterization of transient discharges under atmospheric-pressurenitrogen photoemission and current measurements", J. Phys. D: Appl. Phys. **45** (2012) 125202

[Muller2013] S Müller, T Krähling, D Veza, V Horvatic, C Vadla, J Franzke, "Operation modes of the helium dielectric barrier discharge for soft ionization", Spectrochim. Acta Part B **85** (2013) 104

[Bibinov2001] N K Bibinov, A A Fateev, and K Wiesemann, "Variations of the gas temperature in He/$N_2$ barrier discharges", Plasma Sources Sci. Technol. **10** (2001) 579

[Faure1998] G Faure and S M Shkol'nik, "Determination of rotational and vibrational temperatures in a discharge with liquid non-metallic electrodes in air at atmospheric pressure", J. Phys. D: Appl. Phys. **31** (1998) 1212

[Walsh2010] J L Walsh, F Iza, N B Janson, V J Law, and M G Kong, "Three distinct modes in a cold atmospheric pressure plasma jet", J. Phys. D: Appl.Phys. **43** (2010) 075201

[Moon2002] S Y Moon, W Choe, H S Uhm, Y S Hwang, J J Choi, "Characteristics of an atmospheric microwave-induced plasma generated in ambient air by an argon discharge excited in an open-ended dielectric discharge tube", Phys. Plasmas **9** (2002) 4045

[Wang2007] Q Wang, F Doll, V M Donnelly, D J Economou, N Sadeghi, G F Franz, "Experimental and theoretical study of the effect of gas flow on gas temperature in an atmospheric pressure microplasma", J. Phys. D: Appl. Phys. **40** (2007) 4202

[Guerra2003] V Guerra, P A Sá, J Loureiro, "Electron and metastable kinetics in the nitrogen afterglow", Plasma Sources Sci. Technol. **12** (2003) S8

[Märk1975] T D Märk, "Cross section for single and double ionization of $N_2$ and $O_2$ molecules by electron impact from threshold up to 170 eV", J. Chem. Phys. **63** (1975) 3731

[Fridman2008] A Fridman, "Plasma Chemistry", Cambridge University Press, New York, 2008

[Tozeau1978] M Touzeau and D Pagnon, "Vibrational excitation of $N_2$(C) and $N_2$(B) by metastable argon atoms and the determination of the branching ratio", Chem. Phys. Lett. **53** (1978) 355 [Belikov1997] A E Belikov, "Rotational and vibrational excitation of the $N_2^+$(B) state in a He+N2 electron-beam plasma", Chem. Phys. **215** (1997) 97

[Ono2009] R Ono, C Tobaru, Y Teramoto, and T Oda, "Laser-induced fluorescence of $N_2(A\ ^3\Sigma_u^+)$ metastable in $N_2$ pulsed positive corona discharge", Plasma Sources Sci.Technol. **18** (2009) 025006

[Dyatko2010] N Dyatko and A Napartovich, "Ionization Mechanisms in Ar:$N_2$ Glow Discharge at Elevated Pressures", 41st Plasmadynamics and Lasers Conference (2010) AIAA 2010-4884

[Lindinger1981] W Lindinger, F Howorka, P Lukac, S Kuhn, H Villinger, E Alge, and H Ramler, "Charge transfer of $Ar^+ + N_2 \leftrightarrow N_2^+ + Ar$ at near thermal energies", Phys. Rev. A **23** (1981) 2319

[Märk1971] T D Märk and H J Oskam, "Ion Production and Loss Processes in Helium-Nitrogen Mixtures", Phys. Rev. A **4** (1971) 1445

[Piper1987] L G Piper, "Quenching rate coefficients for $N_2(a'\ ^1\Sigma_u^-)$" J. Chem. Phys. **87** (1987) 1625

[Bibinov2001b] N K Bibinov, A A Fateev, and K Wiesemann, "On the influence of metastable reactions on rotational temperatures in dielectric barrier discharges in He–$N_2$ mixtures", J. Phys. D: Appl. Phys. **34** (2001) 1819

[Hays1973] G N Hays and H J Oskam, "Reaction rate constant for $2N_2(A^3\Sigma_u^+) \rightarrow N_2(C\ ^3\Pi_u) + N_2(X\ ^1\Sigma_g^+, v'>0)$", J. Chem. Phys. **59** (1973) 6088

[Guerra1997] V Guerra and J Loureiro, "Electron and heavy particle kinetics in a low-pressure nitrogen glow discharge", Plasma Sources Sci. Technol. **6** (1997) 361

[Plain1984] A Plain and I Jolly, "Quenching rate constants for $N_2^+(B^2\Sigma_u^+, v' = 0,1,2)$ with $N_2$ and Ne", Chem. Phys. Lett. **111** (1984) 133

[Anicich1993] V G Anicich, "Evaluated Bimolecular Ion-Molecule Gas Phase Kinetics of Positive Ions for Use in Modeling Planetary Atmospheres, Cometary Comae, and Interstellar Clouds", J. Phys. Chem. Ref. Data **22** (1993) 1469

[Brandenburg2005] R Brandenburg, V A Maiorov, Yu B Golubovskii, H-E Wagner, J Behnke, J F Behnke, "Diffuse barrier discharges in nitrogen with small admixtures of oxygen: discharge mechanism and transition to the filamentary regime", J. Phys. D: Appl. Phys. **38** (2005) 2187

[Peñano2012] J Peñano, P Sprangle, B Hafizi, D Gordon, R Fernsler, and M Scully, "Remote lasing in air by recombination and electron impact excitation of molecular nitrogen", J. Appl. Phys. **111** (2012) 033105

[Zheng-De2002] Z-D Kang, Y-K Pu, "Molecular Nitrogen Vibrational Temperature in an Inductively Coupled Plasma", Chin. Phys. Lett. **19** (2002) 211

[Wang2009] C Wang, N Srivastava, S Scherrer, P-R Jang, T S Dibble, Y Duan, "Optical diagnostics of a low power—low gas flow rates atmospheric-pressure argon plasma created by a microwave plasma torch", Plasma Sources Sci. Technol. **18** (2009) 025030

[Motret2000] O Motret, C Hibert, S Pellerin and J M Pouvesle, "Rotational temperature

measurements in atmospheric pulsed dielectric barrier discharge – gas temperature and molecular fraction effects", J. Phys. D: Appl. Phys. 33 (2000) 1493

[Moon2003] S Y Moon, W Choe, "A comparative study of rotational temperatures using diatomic OH, O2 and N2+ molecular spectra emitted from atmospheric plasmas", Spectrochim. Acta Part B 58 (2003) 249

[Bruggeman2014] P J Bruggeman, N Sadeghi, D C Schram and V Linss, "Gas temperature determination from rotational lines in non-equilibrium plasmas: a review", Plasma Sources Sci. Technol. 23 (2014) 023001

[refSpecAir] http://specair-radiation.net/ (last access in May, 2016)

[Humud2015] H R Humud, Q A Abbas, A F Rauuf, "Effect of Gas Flow Rate on The Electron Temperature, Electron Density and Gas temperature for Atmospheric Microwave Plasma Jet", Int. J. Curr. Eng. Technol. **5** (2015) 3819